# Adaptive Intelligent Controller for Household Cooling Systems

Pramit Ghosh[1], Debotosh Bhattacherjee[2] and Soma Datta[3]

[1] Department of Computer Science and Engineering, Academy of Technology, Hooghly-712121, West Bengal, India
pramitghosh2002@yahoo.co.in

[2] Department of Computer Science and Engineering, Jadavpur University, Kolkata, India
debotoshb@hotmail.com

[3] Department of Computer Science and Engineering, Techno India, Salt Lake, Kolkata, India
soma21dec@yahoo.co.in

**Abstract**

*This paper presents a household cooling system controller which is adaptive and intelligent in nature. It is able to control the speed of a household cooling fan or an air conditioner based on the real time data namely room temperature, humidity and time i.e. duration, which are collected from environment. To control the speed in an adaptive and intelligent manner an associative memory neural network has been used. This embedded system is able to learn from training set i.e. the user can teach the system about his/her feelings through training data sets. When the system starts up it allows the fan to run freely i.e. at full speed and after certain interval it takes the environmental parameters like room temperature, humidity and time i.e. duration, as an input and after that system takes the decision and controls the speed of the fan.*

## 1. Introduction

During transition of seasons, that means Summer to Rainy season, autumn to winter, and winter to autumn; the environmental temperature changes very rapidly and people starts feeling cold. When people feel hot then their internal biological body cooling system starts working i.e. sweating, to enhance its effect to feel comfort and people uses some external cooling system like fan, air conditioner.

When environment temperature goes down then the internal body cooling process stops and people feel cold and then people switch off the fan. But when people are in deep sleep most of the time there is no provision to switch of the fan. As a result body temperature goes down and due to that body hormone and enzymes characteristic changes which causes infection in the body. Finally, they have a cold and fever. To protect body from cold an adaptive intelligent controller for household cooling system, that is able to control the speed of the fan, is required.

Air conditioner system [1] is able to keep the room temperature constant but they are costly and consume much power than a fan, these are the major problem in third world countries. Air conditioner works in a closed room [2]. If door or window is opened for long time air conditioning system will not be able to control the room environment, but household cooling fans are free from this type of problem.

To design a controller for this purpose one need to acquire different environmental parameters like temperature, humidity, time etc and a decision is to be taken by fusing them.

## 2. System details

When the cooling system starts up it is allowed to run the fan freely in the full speed, it is called free running mode. After certain interval the system scans the keyboard to know whether the user wants to regulate the speed of the fan via training set or not. If training set is found, then the system decoded the training set. Based on the decoding result system either directly changes the speed of the fan or changes the fan's speed by updating the weight matrix to embed new information. Besides, that system checks the room temperature, humidity, and time, and based on these parameters it sets the speed of the fan. This is shown in fig 1 using a flow diagram. The entire work can be divided into two parts, Hardware circuitry part and associative neural network.

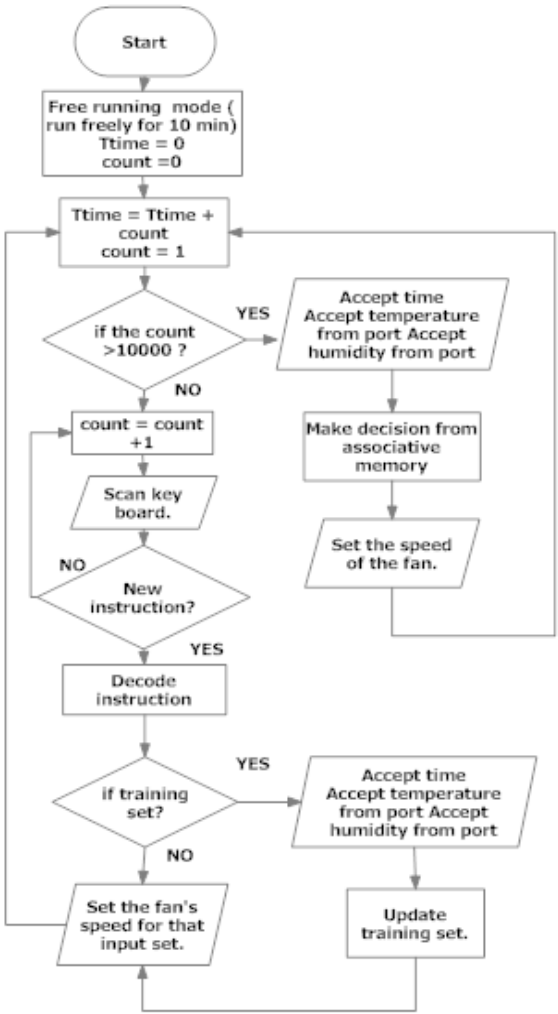

**Fig. 1 The flow diagram of the system.**

## 2.1. Hardware description

For this purpose 8052 [3],[4] based microcontroller namely AT89S52 [5] is used. Because the AT89S52 provides the features: 8K bytes of Flash, 256 bytes of RAM, 32 I/O lines, Watchdog timer, two data pointers, three 16-bit timer/counters, a six-vector two-level interrupt architecture, a full duplex serial port, on-chip oscillator, and clock circuitry. 32 I/O lines are divided into four ports P0, P1, P2, and P3. All these ports are connected with Vcc via 10K resistance, namely pull up resistance, to avoid floating condition, which is shown in fig 2. Fig 2 shows the basic layout of the microcontroller board [6]. In this board there is a provision of 5 volt regulated power supply along with reset button and external crystal connection for clock.

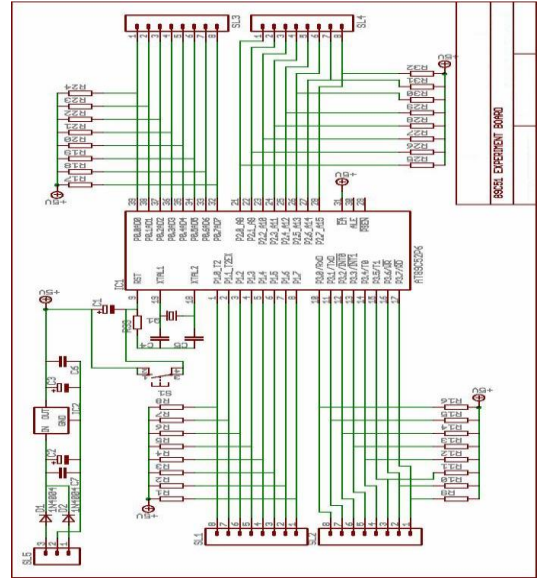

**Fig. 2 Circuit diagram of the 8052 board.**

P1 port is used for keyboard interfacing. Keyboard has 7 valid keys, 0 to 5 this are for fan's speed and one key to inform the system to learn from the training set. P3 and P2 works as an input port shown in fig 3.

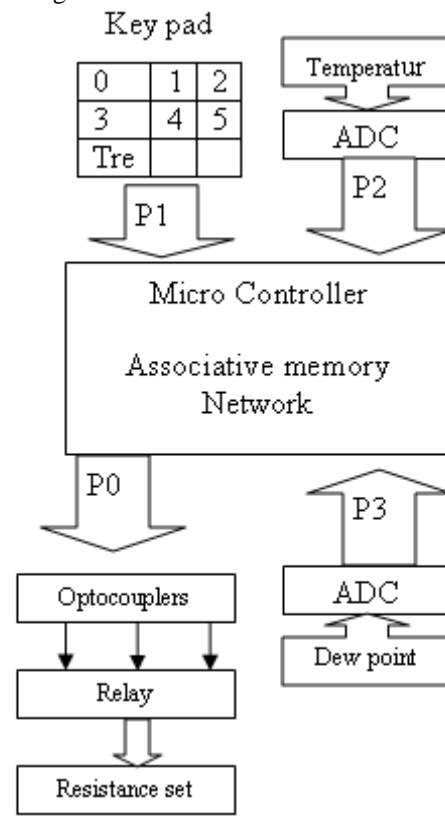

**Fig. 3 Block diagram of the hardware.**

Temperature sensor and dew point sensor are connected with these two ports via respective ADCs [7]. The ADCs namely ADC0804, are configured in free running mode. In this mode pin 18 to pin 11 are connected with either P2 or P3 port to transfer data from ADC. The pin 1, 2,8,10 are connected with ground. Pin 20 is connected with Vcc and ground with a 10uF capacitor to bypass the AC noise as shown in fig 4. P0 is configured as output mode. P0 is used to control the speed of the fan. This P0 is connected with the Optocouplers (MCT2E) for electrical isolation. This MCT2E are used to operate the relays via current amplifiers. Because the current provided by the MCT2E [8] is not sufficient to drive relay. One driving circuit sample is shown in Fig 4. When the relay coil is switched off then a high voltage is generated by the relay coil. To protect the circuit the diode (D1) is used as is shown in Fig 5.

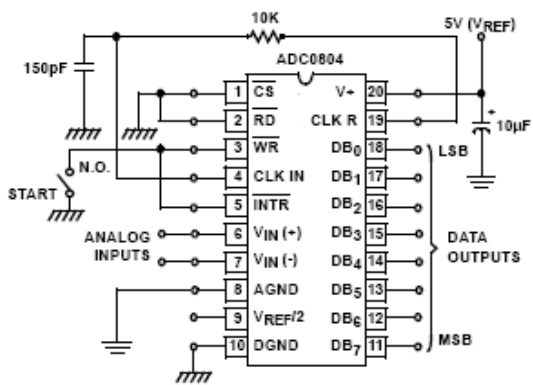

**Fig. 4 ADC is wired in free running mode.**

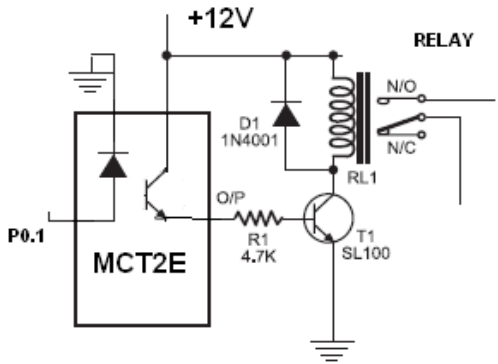

**Fig. 5 Circuit diagram to operate relay.**

## 2.2. Neural network

Associative memory [9] mapping process is used for decision making. However rule based system can make decision but the problem is that no body is able to update the rule during run time. But the associative memory can be trained during runtime. Associative memory network can be seen as a simplified model of a human brain which can associate similar patterns. Associative neural networks are single layer networks in which the weights are determined to store an asset of pattern associations. The associative networks mainly consist of two types of network. If the input vector pair is same as the output vector pair then it results in an auto-associative network. If the input vector pair is different from that of the output vector pair then it forms a hetero-associative network. Hetero-associative network is used in this work for decision making. However, for all the cases, entire process is divided into two parts: Training and Decision making [10].

In training section the weight matrix is created, and in decision section decision is to be taken on the basis of weight matrix. The next two algorithms will explain the training as well as decision making procedure.

**Algorithm1:** Algorithm for Training

The training vector is denoted as s: and testing input vector as x. 'n' is the number of input parameters i.e. size of the input vector, m is the number of output parameters i.e. size of the output vector. 'w' is the weight matrix. $y_j$ is a column vector, contains out put training set in learning process. $x_i$ is a 1×n matrix.

Step1: Initialize all the weights (i=1,.....n, j=1,....m) $w_{ij} = 0$.
Step2 : For each training input and target output vector set, do Step3 to Step5.
Strp3   Set activations for input units to present training input (i=1,2,.......,n),   $x_i = s_i$.
Step4 : Set activations for output units to current target output (j=1,2,.....,m)   $y_j$. $y_j$ is a column vector.
Step5 : Adjust the weights;
$$w_{ij} (new) = w_{ij} (old) + y_j \times x_i \quad (1)$$
Step 6  Stop.

**Algorithm2:** Algorithm for Decision making

W is the weight matrix, x is the input vector.
Step1: Weights ($w_{ij}$) are initialized using Training Algorithm.
Step2 : For each input vector do step3 to step5.
Step3 : Set the activations for input layer units equal to the current vector $x_i$.
Step4 : Compute net input to the output units
$$Y_j = \sum_{i=1}^{n} x_i w_{ij} \quad (2)$$
Step5 : Determine the activation of the output units
$$y_j = \begin{cases} 1,....if\ ..Yj\ =\ \max(\ Yj\ ) \\ 0,....if\ ..Yj\ <\ \max(\ Yj\ ) \end{cases} \quad (3)$$
Step 6  Stop

## 3. Results and Discussions

Table 1 contains the training sets to build the weight matrix. In table 1 "speed of the fan" column contains the speed, column 5 stands for maximum speed, 4 for 2nd maximum speed and so on. Other column contains other input parameters namely temperature, time, and humidity

**Table 1. Training data set.**

| Input parameters | | | Output parameters |
|---|---|---|---|
| Temperature °C | Time in minute (duration) | humidity | Speed of the fan 0,1,2,3,4,5 |
| 39 | 20 | 85 | 0,0,0,0,0,1 |
| 39 | 40 | 85 | 0,0,0,0,0,1 |
| 39 | 60 | 85 | 0,0,0,0,0,1 |
| 38 | 20 | 85 | 0,0,0,0,0,1 |
| 38 | 40 | 85 | 0,0,0,0,0,1 |
| 38 | 60 | 85 | 0,0,0,0,0,1 |
| 36 | 20 | 85 | 0,0,0,0,0,1 |
| 36 | 40 | 85 | 0,0,0,0,0,1 |
| 36 | 60 | 85 | 0,0,0,0,1,0 |
| 34 | 20 | 85 | 0,0,0,0,0,1 |
| 34 | 40 | 85 | 0,0,0,0,1,0 |
| 34 | 60 | 85 | 0,0,0,1,0,0 |
| 29 | 20 | 85 | 0,0,0,0,1,0 |
| 29 | 40 | 85 | 0,0,0,1,0,0 |
| 29 | 60 | 85 | 0,0,1,0,0,0 |
| 28 | 20 | 85 | 0,0,0,0,1,0 |
| 28 | 40 | 85 | 0,0,0,1,0,0 |
| 28 | 60 | 85 | 0,0,1,0,0,0 |
| 27 | 20 | 85 | 0,0,0,0,1,0 |
| 27 | 40 | 85 | 0,0,0,1,0,0 |
| 27 | 60 | 85 | 0,0,1,0,0,0 |
| 26 | 20 | 85 | 0,0,0,0,1,0 |
| 26 | 40 | 85 | 0,0,0,1,0,0 |
| 26 | 60 | 85 | 0,0,1,0,0,0 |
| 25 | 20 | 85 | 0,0,0,0,1,0 |
| 25 | 40 | 85 | 0,0,0,1,0,0 |
| 25 | 60 | 85 | 0,0,1,0,0,0 |
| 23 | 20 | 85 | 0,0,0,0,1,0 |
| 23 | 40 | 85 | 0,1,0,0,0,0 |
| 23 | 60 | 85 | 1,0,0,0,0,0 |
| 22 | 20 | 85 | 0,0,0,1,0,0 |
| 22 | 40 | 85 | 0,1,0,0,0,0 |
| 22 | 60 | 85 | 1,0,0,0,0,0 |
| 21 | 20 | 85 | 0,0,0,1,0,0 |
| 21 | 40 | 85 | 0,1,0,0,0,0 |
| 21 | 60 | 85 | 1,0,0,0,0,0 |
| 20 | 20 | 85 | 0,0,1,0,0,0 |
| 20 | 40 | 85 | 0,1,0,0,0,0 |
| 20 | 60 | 85 | 1,0,0,0,0,0 |
| 19 | 20 | 85 | 0,0,1,0,0,0 |
| 19 | 40 | 85 | 0,1,0,0,0,0 |
| 19 | 60 | 85 | 1,0,0,0,0,0 |
| 18 | 20 | 85 | 0,0,1,0,0,0 |
| 18 | 40 | 85 | 0,1,0,0,0,0 |
| 18 | 60 | 85 | 1,0,0,0,0,0 |
| 17 | 20 | 85 | 0,1,0,0,0,0 |
| 17 | 40 | 85 | 0,1,0,0,0,0 |
| 17 | 60 | 85 | 1,0,0,0,0,0 |

In the training set the humidity is kept as constant and a plot for the training set's temperature, time i.e. duration and value in 3D plane has been computed, which is shown in fig 6. This is because to show human fillings regarding hot environment with respect to temperature and time. The plot reveals that when environment temperature is very near to the human body temperature then the speed of the fan decreases with respect to time increment.

The temperature, time and humidity are shown in decimal format in the Table 1, but in weight matrix calculation they are used in binary format. That's why the Weight matrix has 21 rows (7 × 3) and 6 columns. Fig 7 shows the weight matrix. This weight matrix is created by the "Algorithm for Training" and it will be used by the "Algorithm for Decision making" to make decision.

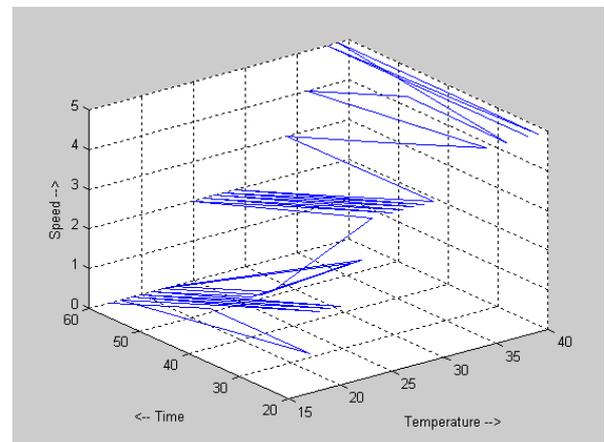

**Fig. 6  3D plot of the temperature, time and speed of the fan.**

MatLab [11] simulation result

For temperature=28° C time = 60 min humidity=85 . test is a matrix to get the value from associative memory.
>> test=getBval(28,60,85);
test = 0 0 1 1 1 0 0 0 0 1 1 1 1 0 1 0 1 0 1 0 1
>>result = test × w   % to find out the value from associative network.
result =  57  32  60  44  42  33
>> t(18,:)  % the result value of training set
ans = 0 0 1 0 0 0

In result '60' is the highest so the decision is ' 0 0 1 0 0 0' which is same as with training set t(18,:). That means 18$^{th}$ row in table 1.

For temperature=40° C time = 10 min humidity=85 which is not available in training set
>> test=getBval(40,10,85);

**test = 0 0 0 1 0 1 0 0 1 0 1 0 0 0 1 0 1 0 1 0 1**

**>> result = test × w % to find out the value from associative network.**

**result = 14  14  20  24  18  28**

In result '28' is the highest so the decision is ' 0 0 0 0 0 1' which is very near to the training set t(1,:)

That means 1st row in table 1.

For temperature=17° C time = 40 min humidity=85

**>> test=getBval(17,40,85);**

**test = 1 0 0 0 1 0 0 0 0 0 1 0 1 0 1 0 1 0 1 0 1**

**>> result = test * w % to find out the value from associative network.**

**result = 50  54  44  46  28  26**

**>> t(47,:) % the result value of training set**

**ans = 0 1 0 0 0 0**

In result '54' is the highest so the decision is ' 0 1 0 0 0 0' which is same as with training set t(47,:)

```
Command Window
Weight matrix
    1    2    0    0    0   -3
    1    0    0    0    0    5
    1    0   -2    0    0    7
   -7   -8    2    2    2   -9
    7    8    8    6    4   -9
   -7   -8   -8   -6   -4    9
   -7   -8   -8   -8   -8   -9
   -7   -8   -8   -8   -8   -9
   -7   -8   -8   -8   -8   -9
    7   -6    8   -2    6    3
    7    6    2    4   -4    1
    7   -6    8   -2    6    3
    7    6    2    4   -4    1
   -7   -8   -8   -8   -8   -9
    7    8    8    8    8    9
   -7   -8   -8   -8   -8   -9
    7    8    8    8    8    9
   -7   -8   -8   -8   -8   -9
    7    8    8    8    8    9
   -7   -8   -8   -8   -8   -9
    7    8    8    8    8    9
```

**Fig. 7  The weight matrix.**

System is able to update its weight matrix set if the training set is given via key board during run time, Table 2 shows the run time training sets.

**Table 2. Training data set during runtime.**

| Input parameters | | | Output parameters |
|---|---|---|---|
| Temperature | Time in minute (duration) | humidity | Speed of the fan 0,1,2,3,4,5 |
| 30 | 20 | 85 | 0,0,0,0,1,0 |
| 30 | 40 | 85 | 0,0,0,1,0,0 |
| 30 | 60 | 85 | 0,0,1,0,0,0 |

The decision making algorithm have been tested for 60 cases, out of which 8 cases does not provides satisfactory results. So the percentage of error is 13.33. This data fusion based controller will be useful for any kind of household cooling system for example air conditioner, air cooler, ceiling mounted fan, table fan, wall mounted fan etc.

## 4. Conclusion

This system is an efficient and cost effective mean to control the speed of a household fan. It is easy to install. It can reduce the probability of cold and fever during the transition season. However, this approach provides 86.67% accuracy but the decision making procedure should be changed to increase the accuracy. To achieve that bidirectional associative memory or back propagation learning algorithm may be helpful.  Future plan of this work is to enhance the learning as well as decision making algorithm. If more training set is used then the decision will be more accurate. As a conclusion it can be said that this Adaptive Intelligent Controller for Household Cooling System has a good market value.

## 5. Acknowledgement